# HARMONIC GENERATION IN A TERAWATT X-RAY FREE-ELECTRON LASER

H.P. Freund[1,2] and P.G. O'Shea[1]

[1]Institute for Research in Electronics and Applied Physics, University of Maryland, College Park, MD 20742, USA
[2]NoVa Physical Science and Simulations, Vienna, Virginia 22182, USA

Terawatt x-ray free-electron lasers (XFELs) require high current densities with strong transverse focusing. The implications on harmonic generation are discussed using the MINERVA simulation code which self-consistently includes harmonic generation. We consider helical and planar undulators where the fundamental is at 1.5 Å and study the associated harmonic generation. While tapered undulators are needed to reach TW powers at the fundamental, the taper does not enhance the harmonics because the taper must start before saturation of the fundamental, with the harmonics saturating earlier. Nevertheless, the harmonics reach substantial powers and enable enhanced applications.

X-ray Free-Electron Lasers (XFELs) are operating or under construction worldwide [1-9] and are attracting an increasing number of users. As such, we expect that novel and important new applications will be found. The Linac Coherent Light Source (LCLS) at the Stanford Linear Accelerator Center [1] was the first XFEL and produced 20 GW pulses of 1.5 Å photons at a repetition rate of 120 Hz, and the other XFELs produce similar powers. The LCLS and many of the other XFELs operate as self-amplified spontaneous emission (SASE) sources; however, recent experiments in self-seeding have shown that it is possible to extract the optical pulse and pass it through a monochromator before reinserting it in phase with the electron beam. There is, of course, interest around the world in methods that produce still higher peak powers.

Recent simulations [10,11] indicate that a terawatt (TW) XFEL is possible. Achieving TW power levels will most likely require a step-tapered, superconducting undulator line; however, it was shown that the most important requirement for a TW XFEL is an extremely strong focusing FODO lattice [11]. In the configurations studied in that work, the current densities reached levels in excess of 1 GA/cm$^2$. Harmonic generation in such XFELs will be dominated by the nonlinear harmonic generation (NHG) mechanism where the harmonics are driven when the fundamental reaches high power levels and the highest harmonic power typically reaches several tenths of a percent of the fundamental power. For a TW XFEL, this implies that the strongest harmonic could reach substantial powers. In the present work, we consider harmonic generation in the configurations studied previously [11] using the MINERVA simulation code [11-15].

MINERVA is a three-dimensional, time-dependent nonlinear formulation for modeling amplifier, oscillator, and self-amplified spontaneous emission (SASE) configurations. MINERVA uses the Slowly-Varying Envelope Approximation (SVEA) and the optical field is represented by a superposition of Gauss-Hermite modes. The field equations are averaged over the rapid sinusoidal time scale and, thereby, reduced to equations describing the evolution of the slowly-varying amplitude and phase. The $x$- and $y$-components of the field are integrated independently; hence, MINERVA is capable of self-consistently simulating an undulator line composed of a variety of different polarizations. Time-dependence is treated using a breakdown of the electron bunch and the optical pulse into temporal *slices* each of which is one wave period in duration. The optical *slices* are allowed to slip ahead of the electron slices. MINERVA integrates each electron and optical *slice* from $z \rightarrow z + \Delta z$ and the appropriate amount of slippage can be applied after each step or after an arbitrary number of steps by interpolation.

Particle orbits are integrated using the full Lorentz force equations in the complete optical and magnetostatic fields (undulators, quadrupoles and dipoles). It is important to remark that the use of the full Lorentz orbit analysis allows MINERVA to self-consistently treat both the entry/exit tapers of undulators, and harmonic generation. In particular, the harmonic treatment has been validated by comparison with a SASE FEL at ENEA Frascati [16] and a tapered amplifier experiment at Brookhaven National Laboratory [17]. Thus, MINERVA self-consistently tracks the particle distribution and optical field through the undulator line and includes optical guiding and diffraction and the associated phase advance of the optical field relative to the electrons throughout.

In this paper, the steady-state simulations discussed previously [11] are extended to include the 2$^{nd}$ through the 5$^{th}$ harmonics. To that end, 248,832 macroparticles are included to ensure convergence. The electron beam parameters correspond to those in the LCLS in which the energy is 13.64 GeV, the current is 4000 A, the emittances are 0.3 mm-mrad in both the $x$- and $y$- directions, and the rms energy spread is 0.01%. Both helical and planar undulators are considered herein with a fundamental resonance at 1.5 Å but the FODO lattice is the same for both undulator lines; specifically, the strong focusing lattice with a cell length of 2.2 m using quadrupoles with a field gradient of 26.4 kG/cm and a length of 7.4 cm [11] corresponding to a beam radius of about 7.1 μm and a current density of 2.5 GA/cm$^2$. The Twiss parameters are chosen to match the electron beam into this lattice. This compares with a current density of about 210 MA/cm$^2$ achieved in the LCLS. Given the strength of the FODO lattice, the beam propagation is only weakly affected by the focusing properties of the planar or helical undulators. Steady-state simulations are adequate since at this wavelength slippage over 100 m of the undulator line is less than 1.5 fs. The LCLS beam is characterized by a flat top temporal profile with a full duration of about 83 fs; hence, slippage is expected to be negligible.

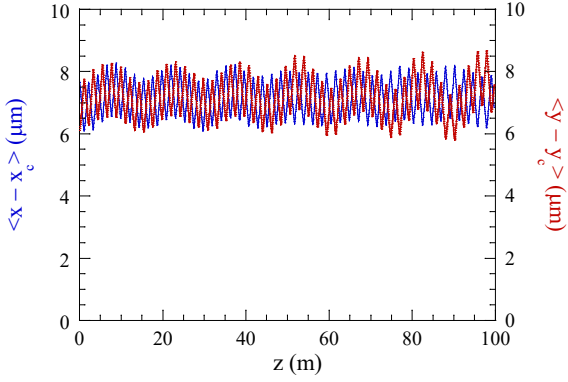

Fig. 1: Evolution of the beam envelope.

Let us first consider the planar undulator line. The evolution of the beam envelope in the *x*- and *y*-directions is shown in Fig. 1. Here we model flat pole face undulators with a maximum on axis field of 12.49 kG, a period of 3.0 cm and a length of 113 periods (3.39 m) with one period entry/exit tapers. The undulator line consists of 25 segments for an overall length, including the drift spaces between the undulators, of 99.7 m. Based on the estimate developed by Ming Xie [18], we expect a saturated power of about 60 GW. A step down-taper was used to enhance the power in the fundamental with a drop of about 0.2% from one undulator to the next. The optimal start taper point differed between the SASE and self-seeded cases.

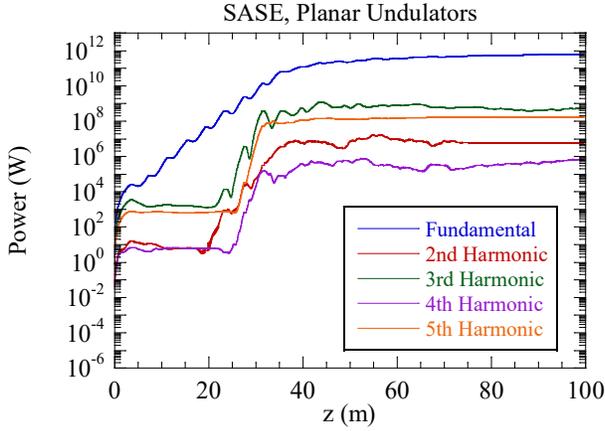

Fig. 2: Evolution of the fundamental and harmonics in the tapered, planar undulator line (SASE).

The start-taper point for the SASE case was found to be the ninth undulator and the fundamental power reached 0.63 TW at the end of the undulator line, which corresponds to an enhancement by a factor of about ten relative to the untapered saturated power. The evolution of the powers at the fundamental and the $2^{nd} - 5^{th}$ harmonics is shown in Fig. 2. It should be noted that this figure corresponds to only a single noise seed. A more complete study would describe an average over at least 15 noise seeds; however, this is enough to show the potential of harmonic generation. It is clear from the figure that (1) the harmonics remain at low power levels until the fundamental power reaches the 1 – 10 MW range and the NHG mechanism becomes effective, and (2) the harmonics are not enhanced by the taper. This is because the harmonics saturate before the fundamental and are not in proper phase for the taper to be effective

A summary of the wavelength, photon energy and output power for the harmonics from the planar undulator line under SASE is shown in Table 1. Observe that, as expected for harmonics from planar undulators, the odd harmonics are more strongly generated than the even harmonics. The most strongly generated harmonic is the $3^{rd}$ and is found at a peak power of 498 MW. The $5^{th}$ harmonic, with a photon energy of 41.33 keV, is much reduced in comparison but is still found at a substantial peak power of 172 MW.

| Harmonic | Wavelength (Å) | Photon Energy (keV) | Power (MW) |
|---|---|---|---|
| 2 | 0.750 | 16.53 | 5.78 |
| 3 | 0.500 | 24.80 | 498 |
| 4 | 0.375 | 33.06 | 0.635 |
| 5 | 0.300 | 41.33 | 172 |

Table 1: Harmonic number, wavelength, photon energy and power in the planar undulator line (SASE).

It is expected that the odd harmonics peak on-axis while the even harmonics peak off-axis, and this is what we find in simulation. However, the NHG mechanism is driven by high fundamental powers which are peaked on-axis and it might be expected that a substantial component of the even harmonic power is also found on-axis, and this is what is found. The dominant modes for the $2^{nd}$ harmonic at the end of the undulator line are the TEM$_{0,0}$ at 0.632 MW, the TEM$_{1,0}$ at 1.73 MW, the TEM$_{2,0}$ at 0.262 MW, the TEM$_{0,2}$ at 0.104 MW, the TEM$_{3,0}$ at 2.33 MW and the TEM$_{1,2}$ at 0.723 MW. Hence, although the dominant modes are peaked off-axis, there is a local maximum on-axis. A similar result is found for the $4^{th}$ harmonic.

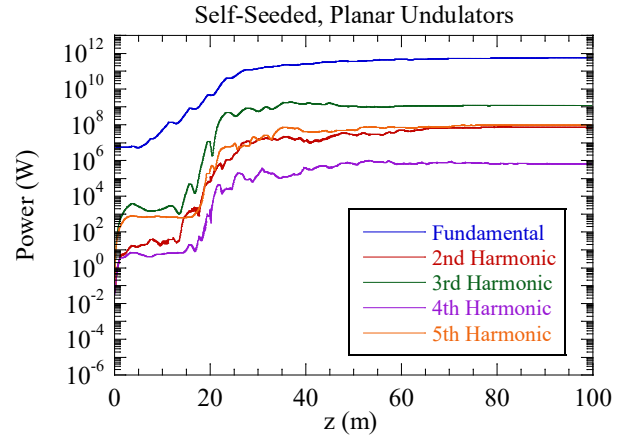

Fig. 3: Evolution of the fundamental and harmonics in the tapered, planar undulator line (self-seeded).

The evolution of the powers at the fundamental and harmonics for the self-seeded planar undulator line is shown in Fig. 3. In this case the fundamental reaches a peak power level of 0.51 TW. While this is somewhat less than that found for the SASE example, it should be noted that the enhancement due to a tapered undulator line is extremely sensitive to the start-taper point and it is difficult to precisely choose the optimal start-taper point in a step-tapered undulator line. Thus, the enhancement will be sensitive to the fluctuations in the SASE power or to the large fluctuations in the self-seeded power.

A summary of the peak powers at the harmonics for the self-seeded example is shown in Table 2. As in the SASE example, the odd harmonics achieve higher powers and the 3rd harmonic power reaches 1.135 GW.

| Harmonic | Power (MW) |
| --- | --- |
| 2 | 64.9 |
| 3 | 1135 |
| 4 | 0.646 |
| 5 | 96.9 |

Table 2: Harmonic powers from the planar undulator line with self-seeding.

Turning to simulations of harmonic generation in helical undulators, it should be noted that the harmonics in helical undulators are generated by a completely different mechanism than is the case for planar undulators [19]. In planar undulators, harmonics are generated due to fluctuations on the axial velocity. In contrast, the axial velocity is constant in helical undulators and harmonics are generated by a phase resonance. The fundamental is found when the polarization rotates through $2\pi$ radians in one undulator period. The $h^{th}$ harmonic is generated when the polarization rotates through $2h\pi$ radians in one undulator period. This emission is into off-axis modes for the incoherent synchrotron radiation and for the linear instability and is generally weaker than is found in planar undulators; however, since it is a phase resonance it is not so sensitive to the magnitude of the undulator field which governs the harmonic generation in planar undulators. Just as in the case of the even harmonics is planar undulators, there is an on-axis component of the harmonics in helical undulators due to the NHG mechanism.

The helical undulators we use have a peak on-axis field of 16.135 kG, a period of 2.0 cm and are 46 periods in length with one period in the entry/exit taper. The optimal down-taper was found to be a drop of 0.08% from segment to segment. The evolution of the powers in the fundamental and harmonics for the SASE example in the step tapered helical undulator line is shown in Fig. 4. It is evident that the fundamental reaches a peak power of 0.93 TW. It is estimated that the saturated power in a uniform undulator line will reach about 82 GW [18], so the taper has resulted in an enhancement by more than a factor of ten. This is greater than for the planar undulator line and is not surprising as the interaction in a helical undulator is typically stronger than in an equivalent planar undulator.

Also, it is seen that, once again, the taper does not enhance the harmonics.

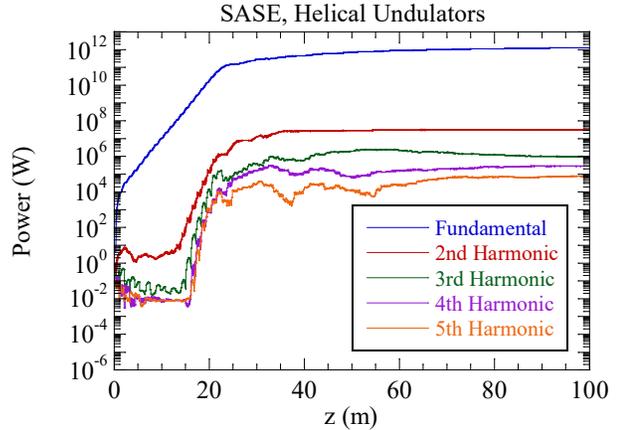

Fig. 4: Evolution of the fundamental and harmonics in the tapered, helical undulator line (SASE).

A summary of the harmonic powers at the harmonics is shown in Table 3. Note that unlike in the planar undulator line the harmonic powers decrease monotonically with increasing harmonic number because there is no distinction between the even and odd harmonics in the phase resonance. Also note that while the fundamental power is higher with the helical undulators, the harmonic powers are reduced relative to the planar undulator line.

| Harmonic | Power (MW) |
| --- | --- |
| 2 | 25.36 |
| 3 | 0.8348 |
| 4 | 0.2606 |
| 5 | 0.0761 |

Table 3: Harmonic powers from the helical undulator line after SASE.

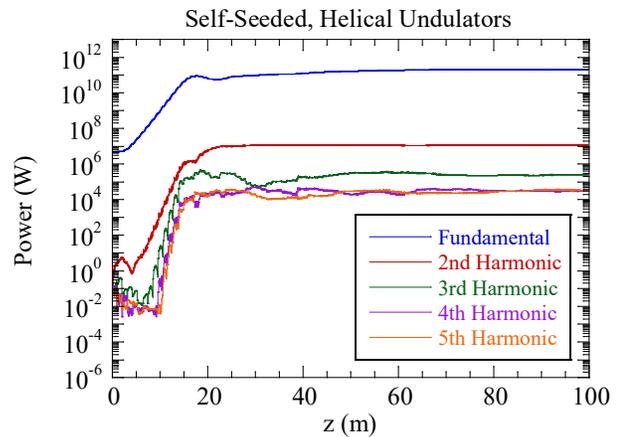

Fig. 5: Evolution of the fundamental and harmonics in the tapered, helical undulator line (self-seeded).

Just as in the case of the even harmonics in the planar undulators, the NHG mechanism is driven by high fundamental powers which are peaked on-axis and there is a substantial component of the harmonic power is also found on-axis. For example, the dominant modes for the 2$^{nd}$ harmonic at the end of the undulator line are the TEM$_{0,0}$ at 2.74 MW, the TEM$_{1,0}$ at 1.68 MW, the TEM$_{0,1}$ at 5.28 MW, the TEM$_{0,2}$ at 5.93 MW, the TEM$_{0,2}$ at 7.04 MW and the TEM$_{0,3}$ at 1.47 MW. Hence, although the dominant modes are peaked off-axis, there is a local maximum on-axis. A similar result is found for the other harmonics.

The evolution of the powers at the fundamental and harmonics for the self-seeded planar undulator line is shown in Fig. 5. In this case the fundamental reaches a peak power level of 0.20 TW. Again, as in the case of the planar undulator line, while this is somewhat less than that found for the SASE example, the reason is that it the step-taper represents a coarse optimization. A summary of the harmonic powers is given in Table 4.

| Harmonic | Power (MW) |
|---|---|
| 2 | 11.26 |
| 3 | 0.244 |
| 4 | 0.0299 |
| 5 | 0.0328 |

Table 4: Harmonic powers from the helical undulator line with self-seeding.

In summary, the development of TW XFELs will give rise to a host of new research opportunities. An earlier work [11], it was demonstrated that terawatt powers in an XFEL requires extremely strong FODO lattices to achieve high current densities. In this paper, the focus has been on the generation of harmonics in these XFELs. Both helical and planar step-tapered undulator lines are studied. It was shown that while the step-taper effectively enhances the fundamental powers, the harmonics saturate before the fundamental and before the optimal start of the step-taper so that the harmonic power is unaffected by the taper. While the helical undulators produce higher fundamental powers than equivalent planar undulators, the planar undulators give rise to higher harmonic powers although strong harmonic generation is found for both configurations. A significant result was that a terawatt XFEL using a planar undulator line is capable of potentially generating hundreds of megawatts of 41 keV photons and substantial powers at still higher harmonics which are much higher than what has been previously achieved at similar wavelengths. These harmonics are tunable across the K-edges of several materials [20]. This opens the possibility of enhanced biomedical K-edge subtraction imaging and x-ray absorption spectroscopy [21].

**ACKNOWLEDGMENTS**

This work was supported by the United States Department of Energy under contract DE-SC0024397.